\documentclass[pra,twocolumn,showpacs,superscriptaddress,floatfix]{revtex4}

\let\a=\alpha \let\b=\beta    \let\d=\delta 
       \let\l=\lambda
\let\m=\mu    \let\n=\nu    \let\x=\xi         \let\r=\rho
\let\s=\sigma    \let\f=\varphi 
     \let\o=\omega
   \let\L=\Lambda 
         
\let\O=\Omega 

\font\tenmib=cmmib10\font\sevenmib=cmmib7\font\fivemib=cmmib5%
\mathchardef\Bl   = "0515  
\newdimen\xshift \newdimen\xwidth \newdimen\yshift \newdimen\ywidth

\def\ins#1#2#3{\vbox to0pt{\kern-#2pt\hbox{\kern#1pt #3}\vss}\nointerlineskip}

\def\eqfig#1#2#3#4#5{
\par\xwidth=#1pt \xshift=\hsize \advance\xshift
by-\xwidth \divide\xshift by 2
\yshift=#2pt \divide\yshift by 2
{\hglue\xshift \vbox to #2pt{\vfil
#3 \includegraphics{#4.eps}
}\hfill\raise\yshift\hbox{#5}}}

\def\V#1{{\bf #1}}
\def\lis#1{{\overline#1}}

\def\RRR{\hbox{\msytw R}}
\def\tende#1{\,\vtop{\ialign{##\crcr\rightarrowfill\crcr
 \noalign{\kern-1pt\nointerlineskip} \hskip3.pt${\scriptstyle
 #1}$\hskip3.pt\crcr}}\,}
\font\titolo=cmbx12%
\font\msytw=msbm10%
\def\iniz{\setcounter{equation}{0}}
\def\be{\begin{equation}}\def\ee{\end{equation}}

\usepackage{fancyhdr}
\begin{document}

\centerline{\titolo 
On Thermostats: Isokinetic }\vskip1mm
\centerline{\titolo or Hamiltonian? finite}\vskip1mm
\centerline{\titolo or infinite?}
{\vskip3mm}

\centerline{Giovanni Gallavotti}
\centerline{Fisica and INFN, Roma1 and IHP, Paris}
\centerline{\today}

{\vskip3mm}
\noindent {\bf Abstract}: {\it The relation between finite isokinetic
thermostats and infinite Hamiltonian thermostats is studied and their
equivalence in the thermodynamic limit is heuristically discussed.}
{\vskip3mm}

\noindent{\bf
Studies on non equilibrium statistical mechanics progressed
considerably after the introduction of artificial forces supposed to
simulate the interaction of a ``test system'' with ``heat
reservoirs'', also called ``thermostats''. Simulations could be
developed eliminating the need of very large systems to model the
action of heat reservoirs.  The drawback is that the equations of
motion are no longer Hamiltonian. The simulations led to developments
and to many new insights into nonequilibrium, particularly with regard
to the theory of large fluctuations (fluctuation theorems, work
relations and attempted applications to problems ranging from
biophysics to fluid turbulence).  An ongoing question has been,
therefore, whether such thermostat models are just devices to generate
simulations that may have little to do with physical reality and,
therefore, in the end not really relevant for Physics. There are,
however, conjectures of equivalence between various kinds of
thermostats and the often preferred ``infinite thermostats'' which,
being Hamiltonian, are considered more fundamental (in spite of being
infinite in size) or the stochastic thermostats. Here we try to
substantiate, via a heuristic analysis, the equivalence conjecture
between ``Hamiltonian'' and ``isokinetic'' thermostats by discussing
it in precise terms. Isokinetic will mean that artificial forces are
introduced whose role is to turn into an exact constant of motion the
total kinetic energy of the particles identified as particles of any
of the thermostats interacting with the test system. The novelty here
is that a careful distinction is made between the test system
particles and the particles of the thermostats in contact with it but
physically located in containers outside the sytem (as in most real
thermal baths). The artificial forces only act on the latter: this is
a substantial difference from most cases considered in the literature
in which the artificial forces act also on the test system particles
(technically called ``bulk thermostats''): it is convenient to call
the thermostats considered here ``peripheral thermostats''. The test
system will be kept fixed but the thermostats will be allowed to be of
arbitrary size, and their behavior as the size becomes infinite is what
will interest us. The conclusion is that, under a suitable assumption,
a peripheral isokinetic thermostat becomes in the thermodynamic limit,
when its container becomes infinite, completely equivalent to a
Hamiltonian infinite thermostat: in the sense that the time evolution
of the configurations ({\it ie} of the phase space point representing
test and interaction systems) is, with probability ${\bf 1}$, the same
as that obtained by letting the isokinetic containers become
infinite. In bulk thermostats there cannot be such strict equivalence
because motion remains non Hamiltonian even in the limit of infinite
systems. The analysis reinforces, as a byproduct, the
identification (modulo an additive total time derivative) between
phase space contraction and entropy production.}

\def\SEC{Thermostats}
\section{Thermostats}\label{sec2}

A classical model for nonequilibrium, for instance in \cite{FV63}, is
a {\it test system} in a container $\O_0$, for instance a sphere of
radius $R_0$ centered at the origin $O$, and several {\it Interaction
systems} containing the {\it thermostats}: we denote their containers
$\O_j$ and they can be thought (to fix ideas) as the sets $\O_j$
consisting of disjoint sectors $\O_j=\{\x\in \RRR^3, |\x|>R_0,
\x\cdot\V k_j<|\x|\,\o_j\}, \,j=1,\ldots,n$, $\V k_j$ distinct unit
vectors, realized, for instance, as disjoint sectors in $\RRR^3$, see
Fig.1, {\it i.e.} as cones in $\RRR^3$ with vertex at the origin
deprived of the points inside the sphere containing the test system:
for precision of language we shall call such containers ``sperically
truncated cones''; but the actual shape could be rather arbitrarily
changed, as it will appear. The terms ``test'' and ``interaction''
systems were introduced in \cite{FV63}. The contact between test
system and thermostats occurs only through the common boundaries
(located on the boundary of the ball $\O_0$) of the test
system. No scaling, of time or space, will be considered here.

In the quoted reference, as well as in later related works,
\cite{ABGM72,Le71,EPR99}, the particles contained in
$\O_0,\ldots,\O_n$ were quantum particles and the interaction systems
were infinitely extended (and obeying a linear Schr\"odinger equation)
and each was initially in a Gibbs state at respective temperatures
$T_1,\ldots, T_n$. Here the particles will be classical, with unit
mass, elastically confined in $\O_0,\O_1\cap\L_r,\ldots,\O_n\cap \L_r$
with $\L_r$ a finite ball, centered at $O$, of radius $r>R_0$. The
temperatures in the interaction systems, here called {\it
thermostats}, will be defined by the total kinetic energies in each of
them: which will be kept a constant of motion by adding a
phenomenological ``thermostatting force''. Hence the qualification of
{\it isokinetic} that will be given to such thermostats. More
appropriately one should call such thermostats ``peripherally
isokinetic'' because most often in the literature the term isokinetic,
instead, refers to systems in which the total kinetic energy of all
particles (in the test and interaction systems) is maintained
constant. The latter are called {\it bulk thermostats}: our models will
correspond to a system in which no internal microscopic friction
occurs and which exchanges energy with external systems kept at
constant temperature. The properties that we discuss cannot hold for
bulk thermostats. However we shall call our thermostats simply
isokinetic except in the last section. The arrangement is illustrated
in Fig.1 below.  \vskip2mm

\noindent {\it Remark:} peripherally isokinetic thermostats have been
considered in the literature in simulations, \cite{LBC92}, and their
physically correct behavior was immediately remarked sparking
investigations about the equivalence problem. See also \cite{ES93,
Ga08}. Recently a case of a model in which only hard core interactions
between particles were present, and the test system was thermostatted
peripherally, has been studied in \cite{GG07} showing the thermostat
action being efficient and measurable even in such extreme situation.

{\vskip3mm} \noindent {\bf Phase space:} {\it Phase space $\cal H$ is
  the collection of locally finite particle configurations $x=(\ldots,
  q_i,\dot q_i,\ldots)_{i=1}^\infty$
\be x=(\V X_0,\dot{\V X}_0, \V X_1,\dot{\V X}_1, \ldots, \V
X_n,\dot{\V X}_n)=(\V X,\dot{\V X})
\label{e1.1}\ee  
with $\V X_j\subset\O_j$, hence $\V X\subset\O=\cup_{j=0}^n \O_j$, and
$\dot q_i\in \RRR^3$; and in every ball ${{\cal B}}(r,O)$, of radius
$r$ and center at the origin $O$, fall a finite number of points of
$\V X$.}  {\vskip3mm}

The space ${{\cal H}}(\L_r)$ will be the
space of the finite configurations with $\V X\subset\L_r$.  It will be
convenient to imagine a configuration $x$ as consisting of a
configuration $(\V X_0,\dot {\V X}_0)\in {{\cal H}}({{\cal
B}}(R_0,O))$ and by $n$ configurations $(\V X_j,\dot{\V X}_j)\in
{{\cal H}}(\O_j\cap \RRR^3/{{\cal
B}}(R_0,O)))$, $j=1,\ldots,n$.

{\vskip3mm} \noindent {\bf Interaction:} {\it The interparticle
interaction $\f$ will be a pair potential with finite range $r_\f$ and
superstable in the sense that $\f$ is non negative, decreasing in its
range (i.e. ``repulsive''), smooth and positive at the origin.}
{\vskip3mm}

\noindent{\it Remark:} Singularities like hard core could be also
considered (at the heuristic level of this paper) but are left out for
brevity. For more general cases, like Lennard-Jones interparticle
potentials or for modeling by external potentials the containers
walls, see \hbox{\cite{BGGZ05}}.  {\vskip3mm}

The potential and kinetic energies of the configuration $x\in{{\cal
H}}(\L_r)$ are $ U(x)=\sum^*_{q',q''\in \V X\cap\L_r} \f(q'-q''),$
$K(x)=\sum_{q_i\in \V X\cap \L_r} \frac{\dot q_i^2}{2}$ where the $*$ means
that the sum is restricted to the pairs $q',q''$ which are either in
the same $\O_j$ or consist of two elements $q',q''$ of which one is in
$\O_0$: this means that particles in $\O_0$ interact with all the
others but the particles in $\O_j$ interact only with the ones in
$\O_j\cup\O_0$. The $\f$'s will be, for simplicity, the same for all
pairs.

{\it The system in $\O_0$ interacts with the thermostats but the
thermostats interact only with the system}, see Fig.1.

\eqfig{240}{86}{%
\ins{38}{20}{
$x=(\V X_0,\dot{\V X}_0,\V X_1,\dot{\V X}_1,\ldots,\V X_n,\dot{\V X}_n)$}
}{fig1}{}

\noindent 
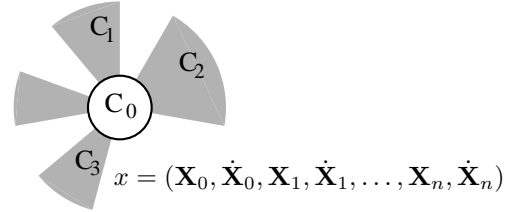
{Fig.1: \small\it The $1+n$ boxes $\O_j\cap\L_r,\, j=0,\ldots,n$,
  are marked $C_0,C_1,\ldots,C_n$ and contain $N_0,N_1,\ldots, N_n$
  particles, mass $m=1$, with positions and velocities denoted $\V
  X_0,\V X_1,\ldots,\V X_n$, and $\dot{\V X}_0,\dot{\V X}_1,\ldots,$
  $\dot{\V X}_n$, respecti\-vely. The $\V E$ are ex\-ter\-nal, po\-si\-tional,
  non conservative, forces; the multipliers $\a_j$ are so defined that
  the kinetic energies $K_j=\frac12 \dot{\V X}_j^2$ are exact
  constants of motion.}  {\vskip3mm}

Hence, if $x\in{{\cal H}}(\L_r)$, the energy $U(x)$ can be written as

\be U(x)=U_0(\V X_0)+\sum_{j=1}^n \big(U_j(\V
X_j)+U_{0,j}(\V X_0,\V X_j)\big)\label{e1.2} \ee
and the kinetic energies will be $K_j(\dot{\V X}_j)=\frac1{2}\dot{\V
X}_j^2$.
The equations of motion will be (see Fig.1)

$$
\ddot{\V X}_{0i}=-\partial_i U_0(\V X_0)-\sum_{j>0}
\partial_i U_{0,j}(\V X_0,\V X_j)+\V E_i(\V X_0)$$
\be \ddot{\V X}_{ji}=-\partial_i U_j(\V X_j)-
\partial_i U_{0,j}(\V X_0,\V X_j)-\a_j \V{{\dot X}}_{ji}\label{e1.3}\ee
where the first label, $j=0,\ldots,n$, denotes the thermostat (or
system) and the second the derivatives with respect to the coordinates
of the points in the correspoding thermostat (hence the labels $i$ in
the subscipts $(j,i)$ have $3N_j$ values); the multipliers $\a_j$ are,
for $j=1,\ldots,n$,

\be \a_j{\,{\buildrel def\over=}\,}\frac{Q_j-\dot U_j}{2K_j},\qquad
Q_j{\,{\buildrel def\over=}\,} -\dot {\V X}_j\cdot \partial_j
U_{0,j}(\V X_0,\V X_j), \label{e1.4} \ee
and the ``walls'' ({\it i.e.} the boundaries
$\partial\O_i,\partial\L_r$) delimiting the different containers will be
supposed elastic. A more general model to which the analysis that
follows also applies is in \cite{Ga06c}.

It is also possible to imagine thermostats acting in the bulk of the
test system by adding a further force $-\a_0 \dot{\V X}_0$: this is,
for instance, of interest in electric conduction models, \cite{Ga96},
where the dissipation is due to energy exchanges with oscillations
(``phonons'') of an underlying lattice of obstacles. Such bulk
thermostatted systems will not be discussed because, for physical
reasons, their dynamics cannot be expected to be equivalent to the
Hamiltonian one in the strong sense that will be considered here. The
thermostat forces would introduce an effective friction on the system
motion not disappearing as the size of the systems grows, as it is
always the case in bulk thermostatted systems.

Other thermostats considered in the literature, \cite{EM90,ECM93},
could be studied and be subject to a similar analysis, which would be
interesting, {\it e.g.} the Nos\'e-Hoover or the isoenergetic
thermostats. Note that even the isoenergetic thermostat does not
conserve Gibbs states (in presence of a test system).

The equations of motion will be called {\it isokinetically
thermostatted} because the multipliers $\a_j$ are so defined to keep
the $K_j$ exactly constant for $j>0$.  The forces $\V E_i(\V X_0)$ are
positional {\it nonconservative}, smooth, forces.  The numbers $N_j$
of particles in the intial data may be random but will be picked with
a distribution giving them average values of $\frac{N_j}{|\O_j\cap
\L_r|}$ within positive and asymptotically $\L_r$-independent bounds
as $r\to\infty$.

{\vskip3mm} \noindent {\bf Initial data:} {\it The probability
  distribution $\m_0$ for the random choice of initial data will be, if
  $dx{\,{\buildrel def\over=}\,}\prod_{j=0}^n\frac{ d\V X_j\,d\dot{\V
      X}_j}{N_j!}$, the limit as $\L_r\to\infty$ of
\be \m_{0,\L_r}(dx)=const\,\, e^{-H_0(x)}
\,dx\label{e1.5} \ee
with $H_0(x)=\sum_{j=0}^n \b_j (K_j(\dot{\V X}_j)-\l_j N_j+
U_j(x))$ and $\b_j{\,{\buildrel def\over=}\,}
\frac1{k_BT_j},\,j>0$ and
$\b_0>0$ arbitrary.}  {\vskip3mm}
Here ${\Bl}=(\l_0,\l_1,\ldots\l_n)$ and ${\bf T}=(T_0,T_1,\ldots,T_n)$ 
are fixed {\it chemical potentials} and {\it temperatures} ($k_B$
being Boltzmann's constant).
\\ 
The limit $\m_0$ as $\L_r\to\infty$ of the distribution in
Eq.(\ref{e1.5}) makes sense (with particles allowed to be located in
the infinite containers $\O_j,\,j>0$) provided it is interpreted as a
Gibbs distribution $\m_0$ obtained by taking the ``termodynamic
limit'' $\L_r\to\infty$, supposing for simplicity that the parameters
$\l_j,T_j,\,j>0$ do not correspond to phase transition points (which
would require care to consider boundary conditions which generate pure
phases, \cite{Ga00}).

It will be convenient to think always the initial data chosen with
respect to the latter distribution: if $\L_r<\infty$ the particles
positions and velocities outside $\L_r$ will, however, be imagined
fixed in time (``frozen'', see \cite{MPP75,CMP000}).
Therefore, defining 

\be Z_j(\l,\b)=\sum_{N=0}^\infty
\int_{(\L_r\cap\O_j)\times\RRR^3} e^{-\b(K_j+U_j-\l N_j)}\frac{ d\V
X\,d\dot{\V X}}{N! }\label{e1.6}\ee
and $\b p_j(\b,\l)=\lim_{\L_r\to\infty}
\frac1{|\L_r\cap\O_j|}\log Z_j(\b,\l)$, the  thermostats density and average
potential energy will be

\be \d_j=\frac{\partial \l_j p_j}{\partial \b_j},\qquad
u_j=-\frac{\partial \b_j p_j}{\partial \b_j} -\frac32k_BT_j\d_j-\l_j\d_j
\label{e1.7}\ee
and $\d_j,u_j, \frac32k_B T_j$ will be suppposed to be the
average density, average potential energy density and average kinetic
energy per particle in the initial configurations: without loss of
generality because this holds {\it with $\m_0$--probability $1$} (by
the no-phase-transitions assumption).

\def\SEC{Dynamics}
\section{Dynamics}\label{sec3}
\iniz

In general time evolution with the thermostatted dynamics changes the
measure of a volume element in phase space by an amount related to
(but different from) the variation of the Liouville volume. 

Minus the change per unit time of a volume element measured via
Eq.(\ref{e1.5}) is, in the sectors of phase space containing $N_j>0$
particles inside $\L_r\cap\O_j$, $j=0,1,\ldots,n$, with kinetic
energy $K_{j,\L_r}(x)$,

\be \s(x)=\sum_{j>0}\frac{Q_j}{k_B
T_j(x)}\,{(1-(3N_j)^{-1})}+\b_0 (\dot K_0+\dot
U_0)\label{e2.1} \ee
where $k_B T_j(x)=\frac23 \frac{K_{j,\L_r}}{N_j}$, and $k_B
T_j(x)\to\b_j^{-1}$ for $\L_r\to\infty$, at least for the initial
data, with $\m_0$ probability $1$.  \vskip1mm

\noindent {\it Remarks:} (1) The dynamics given by the equations of
motion Eq.(\ref{e1.5}) or by the same equations with $\a_j\equiv0$
are of course different. We want to study their difference.

\vskip1mm
\noindent(2) The choice of the initial data with the distribution $\m_0$
regarded as obtained by a thermodynamic limit of Eq.(\ref{e1.5})
rather than (more naturally) with $\m'_{0,\L_r}(dx)$

\be\frac{\m'_{0,\L_r}(dx)}{dx} = const\,\,e^{-H_0(x)} \,\prod_{j=1}^n
\d(K_j-\frac{3 N_j k_B T_j}2)\label{e2.2}\ee
with $N_0,N_1, \ldots,N_n$ fixed, $\frac{N_j}{|\O_j\cap\L_r|}=\d_j$,
$j>0$, and no particles outside $\L_r$ is done to refer, in the
following, to \cite{MPP75,CMP000}. A heuristic analysis would be
possible also with this, and others, alternative choice.

\vskip1mm
\noindent(3)
The Eq.(\ref{e2.2}) is natural, although less convenient notationally,
because in the case $n=1,\V E=\V0$ and $\b_0=\b_1=\b$ with $\b^{-1}=k_B
T_1 (1-\frac1{3N_1})^{-1}$ it is {\it exactly stationary} (a minor
extension of \cite{EM90}), if multiplied by the density
$\r(x)=e^{-\b \sum_{j>0} U(\V X_0,\V X_j)}$, which is the
``missing'' Boltzmann factor in Eq.(\ref{e1.5}), and therefore can be
called an equilibrium distribution.  
\vskip3mm

Choosing initial data with the distribution $\m_{0} $ let $x\to
x^{(\L_r,a)}(t){\buildrel def \over =}
S^{(\L_r,a)}_tx$, $a=0,1$ be the solution of the equations of motion with
$\a_j=0$ ($a=0$, ``Hamiltonian thermostats'') or $\a_j$ given by
Eq.(\ref{e1.3}) ($a=1$, ``isokinetic thermostats'') and ignoring
the particles initially outside $\L_r$, \cite{CMP000}; and
let $S^{(0)}_tx$ be the dynamics $\lim_{\L_r\to\infty} S^{(\L_r,0)}x$.

Existence of a solution to the equations of motion is a problem only
if we wish to study the $\L_r\to\infty$ limit, {{\it i.e.\ }} in the
case in which the thermostats are infinite (thermodynamic limit).

It is a very difficult problem even in the case in which $\a=0$ and
the evolution is Hamiltonian. For $n=1$, $\a_1=0$,
$\b=\b_0{\,{\buildrel def\over=}\,}\b$ and $\V E=\V0$, a case that
will be called {\it equilibrium}, it was shown, \cite{MPP75}, that a
solution to the (Hamiltonian) equations of motion exists for almost
all initial data $x$ chosen with a distribution obtained by
multiplying $\m_{0}(dx)$ by an arbitrary density function $\r(x)$; and
it is defined as the limit as $\L_r\to\infty$ of the finitely many
particles evolutions $S^{(\L_r,0)}_tx$ in $\O\cap\L_r$.

Recently,
the related
problem
of a single infinite system and no thermostat forces 
has been solved in
\cite{MPP75,CMP000} where it has been shown that, for a set of initial
data which have probability $1$ with respect to all distributions like
Eq.(\ref{e1.5}), the Hamiltonian equations make sense and admit a
unique solution, but the general nonequilibrium cases remain open.
\vskip3mm

Therefore in the following I shall suppose, heuristically, a property
(called below ``locality of evolution'') of the equations of motion
Eq.(\ref{e1.3}) {\it with and without} the thermostatting forces
$\a_j\dot {\V X}_j$.  {\vskip3mm}

The question will then be: {\it are the two kinds of thermostats
equivalent?}  {\vskip3mm}

\noindent This is often raised because the isokinetically thermostatted
dynamics is considered ``unphysical'' on grounds that are viewed, by
some, sufficient to ban isokinetic thermostats from use in physically
meaningful problems, like their use to compute transport
coefficients, \cite{EM90}. The following heuristic considerations show that
the latter would be too hasty a conclusion.  

\def\SEC{Heuristic equivalence}
\section{Heuristic discussion and equivalence isokinetic versus
Hamiltonian}\label{sec4}\iniz

The first paper dealing with equivalence issues is \cite{ES93}: its
ideas are taken up here, somewhat modified, and extended. A detailed
comparison with \cite{ES93} is in the last section.

In the Hamiltonian approach the thermostats are infinite systems with
no thermostatting forces ($\a_j\equiv0$) the initial data are still
chosen with the distribution $\m_0$ discussed above.
Let

$$x^{\L_r,1}(t)=(\V X^{\L_r,1}_i(t),\dot{\V
  X}^{\L_r,1}_i(t))_{i=0,\ldots,n}=S^{(\L_r,1)}_t x,$$

\kern-7mm
$$x^{\L_r,0}(t)=(\V X^{\L_r,0}_i(t),\dot{\V X}^{\L_r,i}_0(t))_{i=0,\ldots,n}
=S^{(\L_r,0)}_t x,$$

\kern-7mm
\be x^{0}(t)=(\V X^{0}_i(t),\dot{\V
  X}^{0}_i(t))_{i=0,\ldots,n}=S^{(0)}_t x.
\label{e3.1}\ee
Then a particle $(q_i,\dot q_i)$ located at $t=0$ in, {\it say}, the $j$-th
thermostat evolves, see Eq.(\ref{e1.3}), as

\be q_i(t)=q_i+\int_0^t \dot q_i(t')dt'\hbox{\hglue4cm}\label{e3.2}\ee

\kern-3mm
$$
\dot q_i(t)=e^{-\int_0^t\a_j(t')dt'}\dot q_i+\int_0^t dt''
e^{-\int_{t''}^t \a_j(t')dt'} F_i(t'') dt''$$
where $F_i(t)=-\partial_{q_i} (U_j(\V X_j(t))+U_{j,0}(\V X_0(t),{\V
X}_j(t)))$. 
The above relations hold up to the first collision of the
$i$-th particle with the containers walls, afterwards they hold until
the next collision with a new initial condition given by the elastic
collision rule; they 
hold for the three dynamics considered in
Eq.(\ref{e3.1}) provided $\a_j=0$ in the second and third case and
$\L_r$ is finite in the first and second cases.

The first difficulty with infinite dynamics is to show that the speeds
and the number of particles in a finite region of diameter $r>R_0$
remain finite and bounded in terms of the region diameter (and the
initial data) for all times or, at least, for any prefixed time
interval.

Therefore we shall suppose that the configurations evolve in time
keeping the ``same general statistical properties'' that certainly occur
with probability $1$ with respect to the equilibrium distributions or
the distributions like $\m_0$ in Eq.(\ref{e1.5}): {\it i.e.} density
and velocity that grow at most logarithmically with the size of the
region in which they are observed, \cite{MPP75,CMP000} and average
kinetic energy, average potential energy, average density having,
asyptotically as $\L_r\to\infty$, values $\frac32k_B T_j,\d_j,u_j$
depending only on the thermostats parameters (${\l_j, T_j,\,j>0}$), see
Eq.(\ref{e1.7}).  

More precisely let the {\it local energy} in $\O\cap{\cal B}(\x,R)$,
$\x\in\RRR^3, R> R_0+r_\f$ be

\be W(x;\x,R) =\sum_{q_i\in \V X\cap
{\cal B}(\x,R)}\big(\frac{\dot q_i^2}2+\frac12\sum_{j\ne i}\f(q_i-q_j)+F
r_\f\big)\label{e3.3}\ee
with $F=\max|\partial_q\f|$, and its ``logarithmic scale'' average

\be {\cal E}(x)=\sup_{\x, |\x|>r_\f}
\sup_{R>r_\f\log\frac{2|\x|}{r_\f}} \frac{W(x;\x,R) }{R^3}\label{e3.4}\ee
and call ${\cal H}_0$ the configurations in $\cal H$ with 

$${\cal  E}(x)<\infty \kern3cm {\rm and}$$
\be \lim_{\L_r\to\infty} \frac{N(j,\L_r)}{|\L_r\cap\O_j|}=\d_j,\qquad
\lim_{\L_r\to\infty}
\frac{U(j,\L_r)}{|\L_r\cap\O_j|}=u_j,\label{e3.5}\ee
with $\d_j>0,u_j$ given by Eq.(\ref{e1.7}), if $N(j,\L_r),U(j,\L_r)$
denote the number of particles and their internal potential energy in
$\O_j\cap\L_r)$.  

The set of configurations $x\in{\cal H}_0$ has $\m_{0,\L_r}$-probability
  $1$, \cite{CMP000}.

The discussion in this paper relies on the assumptions 1--3 below,
  motivated by the partial results in \cite{MPP75,CMP000}, as it will
  appear shortly.  It is to be expected that the probability
  distributions $\m^{(\L_r,a,t)},\m^{(0,t)}_t$ obtained by the
  evolution of $\m_0$ with $S^{(\L_r,a)}_t, \, S^{(0)}_t$ ($a=0,1$),
  and all configurations in $x\in{\cal H}_0$ share the following
  properties.

\vskip3mm
\noindent {\bf Local dynamics assumption} {\it 
With $\m_0$-probability $1$ for  for $x\in{{\cal H}_0}$ the number of collisions
$\n(i,t,\L_r,a)$ that
the $i$-th particle of $x^{(\L_r,a)}(t')$ has with the containers walls
for $0\le t'\le t$, is bounded uniformly in $\L_r,a$, and
\\
(1) there is $B(x,t)>0$, continuous and non decreasing in $|t|$,
such that ${\cal E}(x^{(\L_r,a)}(t))$ $\le B(x,t),\, a=0,1$.
\\
(2) The limits $x^{(a)}(t)=\lim_{\L_r\to\infty} x^{(\L_r,a)}(t)$ exist 
and are in ${\cal H}_0$ for all $t$, with ${\cal E}(x^{(0)}(t))\le B(x,t)$.
\\
(3) $x^{(0)}(t)$ 
solve the Hamiltonian equations and the latter  admit a unique solution
in ${\cal H}_0$.
}  \vskip3mm

\noindent{\it Remarks:} (a) The limits of $x^{(\L_r,a)}(t)$, as
    $\L_r\to\infty$, are understood in the sense that for each $i$ the
    limits $(q_i^{(0,a)}(t), p_i^{(0,a)}(t))$ of $(q_i^{(\L_r,a)}(t),
    p_i^{(\L_r,a)}(t))$ exist together with their first two
    derivatives; and $(q_i^{(0,a)}(t), p_i^{(0,a)}(t))$ are twice
    continuously differentiable in $t$ for each $i$. 
    It can be shown that, in the Hamiltonian case $a=0$, the uniform bounds
    in (2) imply the existence of the limits, however they do not
    imply that $x^{(0)}(t)\in{\cal H}_0$, {\it i.e.} they do not imply
    the second of Eq.(\ref{e3.5}).
\\ (b) The number of points of $x^{(\L_r,a)}(t),\, a=0,1$, in a ball
${{\cal B}}(R,\x)$ is bounded by $B(x,t) \,R^3$, for all $R,\x$ with
$R>r_\f\log\frac{2 |\x|}{r_\f}$ and $|t'|<t$.
\\ (c) The speed of a particle located in $q\in\RRR^3$ is bounded by
$B(x,t)(2\log \frac{2|q|}{r_\f})^3$ for $|t'|\le t$.  
\\ (d) Comments (b,c) say that locally the particles keep a finite density and
reasonable energies and momentum distributions. 
\\ (e) An implication is that Eq.(\ref{e3.2}) has a meaning with
probability $1$ on the choice of the initial data $x$. It is very
important that the assumption that dynamics develops within ${\cal
  H}_0$ implies that at all times Eq.(\ref{e3.5}) will hold with
$\d_j,u_j$ {\it time independent}: physically reflecting the infinite
sizes of the thermostats whose density and enegry cannot change in any
finite time.
%
\\
(f) The analysis of the nonequilibrium cases can be partially performed {\it
in similar Hamiltonian cases} as done in the detailed and
constructive analysis in Ref. 16, but dropping
the requirement in Eq.(\ref{e3.5}).
\\(g) It seems reasonable that by the method in \cite{CMP000} the
restriction of satisfying Eq.(\ref{e3.5}) can be removed in the
Hamiltonian model. New ideas seem needed to obtain the local dynamics
property in the case of the thermostatted dynamics.
\vskip3mm
The multipliers $\a_j$ are sums of two terms. The first is
\be \frac{|\dot{\V
X}_j\cdot\partial_j U_{0,j}(\V X_0,\V X_j)|}{ \dot {\V
X}_j^2}\label{e3.6} \ee
see Eq.(\ref{e1.4}) and the short range of the potential implies that
the force $-\partial_j U_{0,j}(\V X_0,\V X_j)$ is a sum of
contributions bounded by $F{\,{\buildrel def\over=}\,} \max|\partial
\f(q)|$ times the number of pairs of particles in the band of width
$r_\f$ around the boundary of the container $\O_0$ (because, by
Eq.(\ref{e3.5}), ${\cal E}(x)<+\infty$: this is of order $O((R_0^2r_\f
F \d)^2)$ if $\d$ is an upper bound on the densities near
$\partial\O_0$. Note that such a bound exists and is time independent,
by the local evolution hypothesis (above), but of course it is not
uniform in the choice of the initial data $x$.

Applying Schwartz' inequality $B_1>0$ exists with:

\be \frac{|\dot{\V
X}_j\cdot\partial_j U_{0,j}(\V X_0,\V X_j)|}{ \dot {\V
X}_j^2}
\le B_1\frac{R_0^2r_\f F \d}{\sqrt {3 N_j k_B T_j \d'}}
\label{e3.7} \ee
for $\L_r$ large and $\d'=\min_{j>0}\d_j$, having used the first of
Eq.(\ref{e3.5}).

The second term in $\a_j$, with $\dot U_j=U(j,\L_r\cap\O_j) $,
contributes to the integrals in the exponentials Eq.(\ref{e3.2}) as
\be {\int_{t'}^t \frac{\dot U_j}{2K_j}dt''}
\simeq \frac{u_j(t)-u_j(t')}{3 k_B T_j}\label{e3.8} \ee
where $u_j(t)$ is the {\it specific} energy at time $t$ and the
$\simeq$ reflects the use of the second equation in Eq.(\ref{e3.6})
to estimate $\frac{U_j}{2K_j}$ as $\frac{U_j}{k_B T_j N(j,\L_r\cap\O_j)}$:
it means equality up to quantities tending to $0$ as $r\to\infty$. 

By the above hypothesis the r.h.s tends to $0$ as $\L_r\to\infty$
because the configurations (initial and after evolution) are in ${\cal
H}_0$, hence have the same specific potential energies $u_j$ (by
Eq.(\ref{e3.5}), see also comment (e) above)), while the contribution
to the argument of the same exponentials from Eq.(\ref{e3.6})
also tends to $1$ by Eq.(\ref{e3.7}).

Taking the limit of Eq.(\ref{e3.2}) {\it at fixed $i$}, this means
that, for initial data in ${\cal H}_0$, hence {\it with
$\m_0$--probability $1$}, the limit motion as $\L_r\to\infty$ (with
$\b_j, \l_j,\,j>0, $ constant) satisfies Hamilton's equations

\be q_i(t)=q_i+\int_0^t \dot q_i(t')dt',\, \dot q_i(t)=\dot
q_i+\int_0^t F_i(t'') dt''\label{e3.9}\ee
and the solution to such equations is unique with probability $1$.

{\it The conclusion is that in the thermodynamic limit the
thermostatted evolution becomes identical, in any prefixed time
interval, to the Hamiltonian evolution on a set of configurations
which have probability $1$ with respect to the initial distribution
$\m_0$, in spite of the non stationarity of the latter.}

In other words. Suppose that the initial data are sampled with the
Gibbs distributions of the thermostats particles (with given chemical
potentials and temperatures) and with an {\it arbitrary distribution}
for the finite system in $\O_0$ (with density with respect to the
Liouville volume, for instance with a Gibbs distribution at
temperature $T_0$ and chemical potential $\l_0$, as in
Eq.(\ref{e1.5})). Then, {\it in the thermodynamic limit}
$\L_r\to\infty$, the time evolution is {\it the same} that would be
obtained, in the same limit, via a isokinetic thermostat acting in
each container $\O_j\cap\L_r$ to keep the total kinetic energy
constant and equal to $\frac32 N_j k_B T_j$.

\def\SEC{Entropy production}
\section{Entropy production}\label{sec5}\iniz

It is important to stress that {\it while, in the thermodynamic limit,
the dynamics becomes the same for isokinetic and Hamiltonian
thermostats, because the thermostat force on each particle tends to
$0$, the phase space contraction in the isokinetic dynamics does not
go to zero}, by Eq.(\ref{e3.7}),(\ref{e3.8}). Instead it becomes, up
to an additive time derivative, see Eq.(\ref{e2.1}),
$\s=\sum_{j>0}\frac{Q_j}{k_B T_j}$. This is possible because $\s$ is a
sum of many quantities (the $\a_j$'s) each of which tends to $0$ in
the thermodynamic limit while their sum does not.

The interest of the remark is that $\sum_{j>0}\frac{Q_j}{k_B T_j}$ is
the natural definition of entropy production in both cases: but in the
literature it is often stated (correctly so in the contexts) that
entropy production is the phase space contraction, raising eyebrows
because the latter vanishes in Hamiltonian models. 

However in finite thermostat models the phase space contraction rate
depends on the metric used to measure volume in phase space: and it
has been stressed that the ambiguity affects the phase space
contraction only by an additive quantity which is a time derivative of
some function on phase space. Such ambiguity will not affect the
fluctuations of the long time averages of the phase space contraction
which, therefore, has an intrinsic physical meaning for this purpose,
\cite{Ga08c}.

  In both the isokinetic and Hamiltonian cases the above $\s$ (which
  is, physically, the physical entropy production) differs, by a time
  derivative $\dot H_{tot}$, from 

\be\lis \s=\sum_{j>0}(\frac{Q_j}{k_B
  T_j}-\frac{Q_j}{k_B T_0}) -\frac{\V E(\V X_0)\cdot \dot{\V X}_0}{k_B
  T_0}.\label{e4.1}\ee
The time derivative in question here is the derivative of the
  total energy $H_{tot}=\b_0(\sum_{j\ge 0}(K_j(x)+U_j(x))+\sum_{j>0}
  U_{0,j}(x))$, \cite{Ga08c}. And $\lis\s$ generates the matter and
  heat currents, \cite{Ga08c}.

  For this reason the equivalence conjectures, of which the
  isokinetic-Hamil\-to\-nian is a prominent example, see \cite{ES93},
  \cite[Sec.8]{GC95b},\cite{Ga95a},\cite[Sec.6]{Ga96},%
  \cite{WKN99},\cite[Sec.9.11]{Ga00}, to quote a few, are relevant for
  the theory of transport and establish a connection between the
  fluctuation dissipation theorem and the fluctuation theorem,
  \cite{Ga08c,Gasch-FL08}.

The works \cite{MPP75,CMP000} bring the present analysis closer to a
mathematical proof for repulsive interaction and I hope to show
in a future work that they actually lead to a full proof of the
locality of the dynamics, at least in dimension $d=1,2$, for other
thermostat models.\def\SEC{Entropy production}
\iniz

\def\SEC{Comparison with \cite{ES93} \& Comments}
\section{Comparison with \cite{ES93} \& Comments}\label{sec6}
\iniz

\noindent (1) Equivalence between different thermostats is widely
studied in the literature and it is {\it surprising} that there are so many
questions still raised about the very foundations, while little
attention is devoted at trying to expand the analysis of the early
works. A clear understanding of the problem was already set up in
comparing isokinetic, isoenergetic and Nos\'e-Hoover bulk thermostats
in \cite{ES93}, where a history of the earlier results is presented as
well, see also \cite{Ru00}.

\noindent (2) Finite thermostats acting on the boundary were studied
already in \cite{LBC92}, in special cases, and were recognized to be
equivalent to thermostats acting on the bulk of the test system. More
recently, \cite{ZRA03}, isokinetic versus isoenergetic thermostats
equivalence has been analyzed and the splitting of the phase space
contraction into an entropy part and an ``irrelevant'' additive time
derivative has been first stressed (see also the later
\cite{GZG05,Ga06c}) and related to the interpretation and prediction
of numerical simulations.

\noindent (3) The basic idea in \cite{ES93} for the equivalence is
that the multipliers defining the forces that remove the heat in
finite thermostat models have equal average (``equal dissipation'') in
the thermodynamic limit, \cite[Eq.(15)]{ES93}: thus making all
evolutions equivalent. In \cite{ES93} the expectation of observables
in two thermostatted evolutions is represented via Dyson's expansion
of the respective Liouville operators starting from an equilibrium
distribution: equivalence follows order by order in the expansion (in
the joint thermodynamic limit and infinite time limit) if a mixing
property, \cite[Eq.(23)]{ES93}, of the evolution with respect to {\it
both} the equilibrium and the stationary distributions is assumed. The
method is particularly suitable for bulk thermostatted systems close
to equilibrium where application of Dyson's expansion can be
justified, at least in some cases, \cite{CELS93}.

\noindent (4) The main difference between the present work and
\cite{ES93} is that here, even far out of equilibrium, we discuss
equivalence between the boundary thermostatted dynamics and
Hamiltonian dynamics: therefore we compare a situation in which the
average value of the dissipation (analogue of \cite[Eq.(32)]{ES93}) is
$\ne0$ with one in which it is $0$ exactly, at least formally. {\it
\\
This is achieved by showing that the multipliers in the models in
Fig.1 vanish in the thermodynamic limit not only in average but also
pointwise with probability $1$}; this is in agreement with the results
in \cite{LBC92} and provides more theoretical grounds to explain
them.
\\
It also means that in boundary thermostatted systems the
analogue of \cite[Eq.(32)]{ES93} does not tend to $0$ when
$N\to\infty$ although the analogue of the average of the multipliers,
corresponding to \cite[Eq.(33)]{ES93}, does.

\noindent (5) In bulk thermostatted systems there {\it cannot be
equivalence between the Hamiltonian and the isokinetic dynamics} in
the sense discussed in this paper, {{\it i.e.}} identity of the
dynamics of individual particles. However, as discussed already in
\cite{ES93}, the expectation values of extensive observables could
hold. On the other hand the analysis in \cite{ES93} should be
extendible to cover also the boundary thermostatted systems because,
while the dissipation ({{\it i.e.}}  entropy production) does not
vanish in the thermodynamic limit, the average of the multipliers
still does, see (3) above, and this is what is really needed in
\cite{ES93}.

\noindent (6) Neither Dyson's expansion convergence questions nor
time-mixing properties, on which \cite{ES93} is based, enter into the
present analysis: but the assumptions needed on the dynamics (local
dynamics) are still strong and are only under partial control via the
theory in \cite{MPP75,CMP000}.

\noindent (7) An important question is whether taking the time
$t\to\infty$ limit after the thermodynamic limit $\L_r\to\infty$
(when, therefore, the dynamics are identical) the probability
distribution $S^{(0)}_t\m_0$ tends to a limit $\m$, and $\m$
still attributes probability $1$ to ${\cal H}_0$: this is an
apparently much harder question related to the difference between the
transient results and the, deeper, steady state results, \cite{CG99}.

\noindent (8) Finally: the choice, made here, of dimension $3$ for the
ambient space is not necessary for the analysis. Dimension $d=1,2,3$
would be equally suited. However it is only if the thermostats
containers dimension is $d=3$ that the system
with infinite thermostats is expected to reach a stationary state: if
$d=1,2$ the equalization of the temperatures is expected to spread
from the system to the reservoirs and proceed indefinitely tending to
establish a constant temperature over larger and larger regions of
size growing with a power of time, \cite{Ru98}.

\vskip3mm

\noindent {\bf Acknowledgements:} I am grateful to C. Maes for
bringing up the problem, and to him, E. Presutti and F. Zamponi for
criticism and hints: E. Presutti made the essential suggestion to use
\cite{CMP000} to try to put the ideas in a precise mathematical
context and to state properly the conditions on the walls
collisions. I am also grateful to the referees for their remarks that
I have incorporated in the revised version. Work partially supported
by I.H.P., Paris, at the workshop {\it Interacting particle systems,
statistical mechanics and probability theory}, 2008, through the {\it
Fondation Sciences Math\'ematiques de Paris}.


\small \bibliographystyle{unsrt}

\end{document}